\documentclass[twocolumn, switch]{article} 

\usepackage{preprint}

\usepackage{amsmath, amsthm, amssymb, amsfonts}

\usepackage[numbers,square]{natbib}
\usepackage[T1]{fontenc}	
\usepackage{xcolor}		
\usepackage[colorlinks = true,
            linkcolor = purple,
            urlcolor  = blue,
            citecolor = cyan,
            anchorcolor = black]{hyperref}	
\usepackage{booktabs} 		
\usepackage{array}			
\usepackage{tabularx}		
\usepackage{multirow}		
\usepackage{threeparttable}	
\usepackage{caption}		
\usepackage{newunicodechar}	
\usepackage{nicefrac}		
\usepackage{microtype}		
\usepackage{lineno}		
\usepackage{float}			

\newcolumntype{Y}{>{\raggedright\arraybackslash}X}
\newcolumntype{T}{>{\raggedright\arraybackslash}X}
\newcolumntype{P}[1]{>{\raggedright\arraybackslash}p{#1}}
\newcolumntype{C}[1]{>{\centering\arraybackslash}p{#1}}
\newcommand{\piilist}[1]{\begingroup\raggedright\arraybackslash #1\endgroup}
\providecommand{\Description}[1]{}
\makeatletter
\newcommand{\newunicodecharifneeded}[2]{%
  \ifcsname u8:\detokenize{#1}\endcsname
  \else
    \newunicodechar{#1}{#2}%
  \fi
}
\makeatother
\newunicodecharifneeded{’}{'}
\newunicodecharifneeded{–}{--}
\newunicodecharifneeded{—}{---}
\newunicodecharifneeded{š}{\v{s}}
\newunicodecharifneeded{č}{\v{c}}
\newunicodecharifneeded{ć}{\'{c}}
\newunicodecharifneeded{ř}{\v{r}}
\newunicodecharifneeded{ž}{\v{z}}
\newunicodecharifneeded{ş}{\c{s}}
\newunicodecharifneeded{ł}{l}
\newunicodecharifneeded{•}{\textbullet}
\makeatletter
\g@addto@macro\UrlBreaks{\do\-\do\_\do\.\do\?\do\=\do\&}
\makeatother

\usepackage{lipsum}		

\usepackage{newfloat}
\DeclareFloatingEnvironment[name={Supplementary Figure}]{suppfigure}
\usepackage{sidecap}
\sidecaptionvpos{figure}{c}

\usepackage{titlesec}
\titlespacing\section{0pt}{12pt plus 3pt minus 3pt}{1pt plus 1pt minus 1pt}
\titlespacing\subsection{0pt}{10pt plus 3pt minus 3pt}{1pt plus 1pt minus 1pt}
\titlespacing\subsubsection{0pt}{8pt plus 3pt minus 3pt}{1pt plus 1pt minus 1pt}

\usepackage{tikz,xcolor,hyperref}

\definecolor{lime}{HTML}{A6CE39}
\DeclareRobustCommand{\orcidicon}{
	\begin{tikzpicture}
	\draw[lime, fill=lime] (0,0)
	circle [radius=0.16]
	node[white] {{\fontfamily{qag}\selectfont \tiny ID}};
	\draw[white, fill=white] (-0.0625,0.095)
	circle [radius=0.007];
	\end{tikzpicture}
	\hspace{-2mm}
}
\foreach \x in {A, ..., Z}{\expandafter\xdef\csname orcid\x\endcsname{\noexpand\href{https://orcid.org/\csname orcidauthor\x\endcsname}
			{\noexpand\orcidicon}}
}

\title{Enabling Multilingual Privacy Policy Audits:\texorpdfstring{\\}{: }Large-Scale Analysis of Spanish Mobile Apps}

\usepackage{authblk}

\author[1]{Marcos Moran\orcidA{}}
\author[2]{David Rodriguez\orcidB{}}
\author[3]{Luka Nenadic\orcidC{}}
\author[4]{Norman Sadeh\orcidD{}}
\author[5]{Jose M. Del Alamo\orcidE{}}

\affil[1, 2, 5]{Universidad Politécnica de Madrid}
\affil[2, 5]{Information Processing and Telecommunications Center}
\affil[3]{ETH Zurich}
\affil[4]{Carnegie Mellon University}

\begin{document}

\twocolumn[ 
  \begin{@twocolumnfalse} 

\maketitle

\begin{abstract}
Automated analyses of privacy policies enable large-scale assessments of transparency in digital ecosystems, yet existing auditing pipelines remain predominantly English-centric. This limits their ability to systematically evaluate multilingual environments, as in the European Union, where many services disclose privacy practices only in local languages. This paper examines whether large language models (LLMs) can extend privacy policy analysis beyond English without requiring language-specific adaptation, thus empowering large-scale auditing in linguistically diverse app ecosystems.

We assemble an evaluation corpus spanning all 24 official EU languages from translated versions of two established expert-annotated datasets (OPP-115 and MAPP) and assess translation fidelity through automated metrics and targeted legal-expert review. Our LLM-based classifier for identifying categories of personal data collection achieves stable cross-lingual performance, with macro-F1 scores ranging between 0.91 and 0.94.

We then leverage this capability in a large-scale audit of 2,611 Android applications from the Spanish Google Play Store. Combining multilingual privacy policy analysis with the evaluation of corresponding privacy labels and runtime network traffic exposes an important linguistic barrier: public-sector apps predominantly provide privacy policies in Spanish, whereas popular commercial apps mostly provide them in English. We reveal systematic discrepancies between declared and observed practices, especially in public-sector apps. Overall, our results indicate how English-only privacy audits can systematically obfuscate transparency gaps in multilingual environments.
\end{abstract}
\vspace{0.35cm}

  \end{@twocolumnfalse} 
] 



\section{Introduction}
Transparency is a foundational obligation of modern data protection law. Regulations such as the EU's General Data Protection Regulation (GDPR)~\cite{gdpr2016}, the California Consumer Privacy Act (CCPA)~\cite{ccpa2018}, and Brazil's Lei Geral de Proteção de Dados (LGPD)~\cite{lgpd2018} require organizations to disclose, in an intelligible and accessible manner, how personal data are collected, used, and shared. On app stores, privacy policies remain the principal channel for these disclosures, while privacy labels~\cite{kelley2009nutritionlabel,kelley2010standardizing} provide a complementary and more accessible format. 

However, unlike platform-mediated labels, privacy policies are often linked as external documents and may be exclusively available in a non-English language. In the European Union, with 24 official languages~\cite{europeancommission_languages} and uneven English proficiency~\cite{eurobarometer2012languages,terstegge_gdpr_lost_in_translation}, policy intelligibility is inherently a multilingual problem. Research and oversight methods that can only process English privacy policies cannot assess transparency uniformly across the EU.

Given the scale and complexity of modern digital ecosystems, systematic scrutiny of privacy disclosures increasingly relies on automated analysis~\cite{amos2021privacypolicies,wagner2023privacy,mhaidli2023researchers}. Yet the tools and benchmarks that enable such scrutiny have been developed and validated predominantly in English. Earlier NLP and machine-learning approaches---from rule-based systems to neural architectures such as Polisis~\cite{harkous2018polisis} and MAPS~\cite{zimmeck2019maps}---rely on language-specific annotated datasets that are difficult to reproduce across languages, which has mostly confined large-scale privacy audits to English-language corpora~\cite{delAlamo2022systematic}. This monolingual bias acts as a structural filter on which parts of the digital ecosystem can be assessed systematically and which remain far less scrutinized. Applications with privacy policies in local or regional languages---including many public-sector services, whose disclosures often follow the language of the jurisdiction they serve---are, therefore, less likely to be captured by the analyses that researchers and regulators rely on to detect transparency gaps.

Recent advances in large language models (LLMs) offer a promising direction for cross-lingual privacy policy analysis. Unlike earlier supervised approaches, LLMs support zero- and few-shot classification without language-specific fine-tuning~\cite{tang2023policygpt,rodriguez2024largeLanguageModelsPrivacyAnalysis}, potentially reducing the annotation burden that has historically constrained multilingual evaluation. However, assessing whether these capabilities transfer reliably to privacy policy analysis still requires annotated reference data in the target languages. Constructing such resources across all 24 official EU languages would require legal experts with both domain-specific knowledge and sufficient proficiency in each language; a combination that is difficult to obtain at scale. To the best of our knowledge, no broadly multilingual annotated corpus currently exists for privacy-policy analysis. Consequently, it remains unclear whether LLM-based methods can support reliable transparency audits involving non-English policies.

This paper thus investigates (1) whether LLMs can reliably assess policies beyond the English language; and (2) how this capability could support empirical auditing in multilingual app ecosystems. To this end, we construct a derived multilingual evaluation corpus covering all 24 official EU languages by translating the expert-annotated OPP-115~\cite{wilson2016creation} and MAPP~\cite{arora2022corpus} corpora using the European Commission's eTranslation service~\cite{eu_translation_ai_tools,arnejsek2020multidimensional}. We retain the source annotations for cross-lingual consistency and assess translation fidelity through automated metrics and targeted legal-expert sanity checks. We then evaluate an LLM-based method~\cite{rodriguez2024largeLanguageModelsPrivacyAnalysis} for identifying categories of personal data collection in policies according to a predefined taxonomy. On this derived, machine-translated benchmark, the method achieves stable performance across languages, with macro-F1 scores ranging from 0.91 to 0.94.

We operationalize this capability in a large-scale audit of 2,611 Android applications from the Spanish Google Play Store, combining multilingual privacy policy analysis with privacy labels and runtime network traffic. This setting allows us to examine how the language of privacy policies affects audit coverage: we find that public-sector apps predominantly provide privacy policies in Spanish, whereas popular commercial apps mostly offer disclosures in English. The audit includes applications that would otherwise remain outside English-centric analysis and reveals systematic discrepancies between declared and observed data practices. These discrepancies are especially pronounced among public-sector apps and often involve device and system telemetry transmitted through SDKs, cloud services, platform APIs, and hosted infrastructure.

\textbf{Contributions.} This paper makes the following contributions:
\begin{enumerate}
    \item \textit{Derived Multilingual Evaluation Corpus.} We construct and release\footnote{The translated multilingual corpora will be moved to a public repository upon the paper's acceptance.} a corpus covering all 24 official EU languages by translating OPP-115 and MAPP while preserving their expert annotations, enabling controlled cross-lingual evaluation methods.

    \item \textit{Cross-lingual LLM Evaluation.} We evaluate an LLM-based method for identifying categories of personal data collection across this corpus without language-specific adaptation, observing stable performance across languages on the derived benchmark.

    \item \textit{Large-scale Empirical Audit.} We apply the method to 2,611 Android apps from the Spanish Google Play Store, jointly analyzing privacy policies in Spanish and English, privacy labels, and runtime network traces to study how the language of privacy policies affects audit coverage and disclosure gaps across public-sector and commercial apps.
\end{enumerate}

\section{Related Work} \label{sec:related_work}

\paragraph{Privacy Policy Corpora and Benchmarks.}
Large collections~\cite{srinath2021privacy} and longitudinal datasets~\cite{wagner2023privacy} provide broad access to privacy policy texts but lack expert practice-level annotations, limiting their utility as supervised benchmarks. Expert-labeled datasets, by contrast, capture richer semantics at the expense of breadth. OPP-115~\cite{wilson2016creation} introduced structured, segment-level annotations across ten privacy practices in 115 website privacy policies and remains the canonical benchmark for policy understanding. APP-350~\cite{zimmeck2019maps} includes 350 Android app privacy policies manually labeled for 18 practices. Other studies relate to cross-border data transfer disclosures~\cite{guaman2023automatedGDPRCompliance} and legitimate-policy detection~\cite{boldt2019analysis}. Although these works established valuable baselines for automated privacy policy analysis, their predominantly monolingual designs limit systematic assessment of multilingual methods and real-world environments.

\paragraph{Multilingual Resources and Cross-Lingual Transfers.}
Efforts to address multilingualism remain limited and fragmented. The Privacy Law Corpus corpus~\cite{gupta2024internationalcorpusofprivacylaws} aggregates privacy laws and guidelines from 183 jurisdictions in 54 languages, with most entries relying on machine translation. In adjacent legal-text domains, Terms of Service (ToS) have supported analysis of potentially unfair clauses~\cite{lippi2019claudette}, including across selected European languages~\cite{drawzeski2021corpus}. Bernhard et al.~\cite{bernhard2025multilingualScraper} released a scraper for large-scale multilingual collection of privacy and ToS documents, but the resulting corpus lacks semantic practice annotations. MAPP~\cite{arora2022corpus} introduced bilingual annotations over 59 English- and German-language app-policy pairs. However, only nine of these pairs were near-identical, and seven of those included separate sections for EU users, limiting their utility for evaluating cross-lingual consistency. More recently, Nenadic et al.~\cite{nenadic2025automatedBoilerplate} introduced a multilingual dataset of 120 policies in English, German, French, and Italian annotated for GDPR- and Swiss-law disclosures. These efforts show growing interest in multilingual resources, but they do not provide a comprehensive multilingual benchmark with comparable labels across all EU languages.

\paragraph{Automated Analysis: From Rules to Deep Learning.}
Automated privacy-policy analysis has evolved from rule-based and classical machine-learning methods to neural architectures. Ontology-driven frameworks such as PrivOnto~\cite{vilaza2018privonto} represented annotated privacy practices, while PolicyLint~\cite{policyLint2019} encoded data-collection and sharing statements as actor--action--data object--entity tuples to detect internal contradictions. Other methods combined heuristics and supervised learning to identify opt-out choices~\cite{kumar2020findingChoiceHaystack}, or used syntactic patterns to extract hyponymy relations among information types~\cite{evans2017evaluationConstituencyHyponymy}. OPP-115 and APP-350 enabled evaluating supervised segment-level classification methods at scale~\cite{wilson2018analyzingPrivacyPoliciesAtScale,zimmeck2019maps}, followed by neural models for vagueness detection and structured policy querying, including BiLSTMs and Polisis~\cite{lebanoff2018automaticDetectionVaguePrivacyPolicies,harkous2018polisis}. Cross-language work remains rare, such as bilingual efforts for detecting privacy and cookie policies and longitudinal English--German analyses of regulatory impact~\cite{hosseini2021unifyingPrivacyPolicyDetection,hosseini2024bilingualLongitudinal}. Overall, most approaches remain language-specific and data-intensive, requiring labeled data or retraining for each language or domain.

\paragraph{LLM-based Privacy Auditing.}
Recent advances in LLMs have introduced new opportunities for privacy-policy analysis by supporting efficient annotation~\cite{cevallos2025gpt} and automated extraction of data practices from policy texts. Tang et al.~\cite{tang2023policygpt} showed that GPT-3.5 can predict high-level data-practice categories in OPP-115 without task-specific fine-tuning, while Rodriguez et al.~\cite{rodriguez2024largeLanguageModelsPrivacyAnalysis} reported high F1 scores for a prompt-engineered GPT-4 Turbo configuration in policy-level extraction of data practices and compared its performance with Llama 2. Nenadic et al.~\cite{nenadic2025automatedBoilerplate} further explored the prevalence of multilingual transparency disclosures required by Swiss and EU regulations using GPT models. Existing evaluations, however, cover only a limited number of languages.

\paragraph{Illuminating the Blind Spots.}
Our study addresses the gaps in the literature by deriving translated versions of OPP-115 and MAPP across all 24 official EU languages and evaluating LLM-based classification of stated data practices across them. We then use this capability in a mobile privacy audit that combines privacy policy analysis, app-store labels, and runtime network observations.

\section{Generating Multilingual Corpora}
\label{sec:multilingual_corpora}

Evaluating multilingual privacy-policy analysis requires a benchmark in which labels are comparable across languages. We, therefore, construct a derived multilingual extension of two expert-annotated English corpora: OPP-115~\cite{wilson2016creation} and MAPP~\cite{arora2022corpus}. Using both corpora reflects the heterogeneous disclosures found in website and app-store audits. In the latter case, apps may link to broader service-level privacy policies rather than documents written exclusively for the mobile app~\cite{rodriguez2024largeLanguageModelsPrivacyAnalysis}.

We preserve the original expert labels and translate the underlying English texts into the 24 official EU languages. This design enables controlled cross-lingual evaluation under comparable annotation conditions, since the semantic labels remain aligned across translated versions of the same source policies.

We first compare GPT-4o and eTranslation as candidate translation systems (\S\ref{subsec:translation_tools}--\S\ref{subsec:translation_validation}). We then use the selected system to generate the multilingual corpora (\S\ref{subsec:dataset_translations}) and assess translation fidelity through targeted legal-expert review (\S\ref{subsec:manual_validation}).

\subsection{Translation Tools and Metrics}
\label{subsec:translation_tools}

To produce multilingual versions of the annotated privacy policy corpora, we compared two translation systems: GPT-4o and eTranslation. GPT-4o is an LLM with strong multilingual capabilities and competitive translation performance across diverse domains~\cite{openai2024helloGPT4o}. eTranslation is an official translation service of the EU, optimized for legal and administrative texts and supporting all 24 official EU languages~\cite{eu_translation_ai_tools}. Its use for translating EU laws and other legal documents makes it particularly relevant for our context.

The quality of the two models' translations was evaluated using four complementary metrics: BLEU~\cite{papineni2002bleu}, ChrF~\cite{popovic2015chrf}, METEOR~\cite{banerjee2005meteor}, and COMET-22 (hereafter COMET)~\cite{rei2020comet}. BLEU measures n-gram overlap between reference and candidate translations, whereas ChrF operates at the character level and better captures inflectional variation in morphologically rich languages. METEOR extends these surface metrics by accounting for synonymy, stemming, and word order, yielding higher correlation with human judgments than earlier lexical metrics~\cite{machacek2013wmt13metrics, machacek2014wmt14metrics}. COMET leverages multilingual pretrained encoders fine-tuned on human adequacy assessments, achieving state-of-the-art correlation with expert ratings~\cite{mathur2020wmt20metrics, freitag2022wmt22metrics}.

Because these metrics do not provide corpus-independent thresholds for legal or policy translation quality, we use them comparatively rather than as absolute guarantees of fidelity. Agreement across metrics increases confidence in system-level trends, while divergences help identify language- or domain-specific sensitivities that require closer inspection.

\subsection{Evaluation of Translation Tools}
\label{subsec:translation_validation}

\subsubsection{Reference Corpora}
\label{subsubsec:reference_corpora}

Evaluating the translation tools' quality requires reference corpora with reliable parallel translations. The main challenge lies in identifying one that is both linguistically robust across all EU languages as well as independent from the training data of the tested translation tool---two requirements that are difficult to satisfy simultaneously.

Most multilingual legal corpora originate from European institutions. EUR-Lex~\cite{eurlex} provides official EU treaties, directives, and case law in 24 languages but is aligned primarily at the article level, offering limited parallelism. CELLAR~\cite{cellarPortal} offers legal texts too but suffers from structural heterogeneity and inconsistent alignment. JRC-Acquis ~\cite{steinberger2006jrcacquis} offers the most suitable benchmark, providing sentence-aligned translations of official legal texts in 22 EU languages (excluding Irish and Croatian) with consistent formatting and legal fidelity. For our evaluation, we extracted 77 documents related to data protection from JRC-Acquis.

Using EU institutional corpora introduces a potential threat to validity due to shared provenance with the data used to train eTranslation. While the degree of direct overlap between our evaluation subset and the system’s training material is unknown, partial exposure to related documents is plausible. Such domain familiarity could advantage eTranslation on EU legal content, whereas GPT-4o’s exposure, if any, would be indirect and diluted across general-domain web data. To assess whether performance differences arise from this effect or reflect general translation quality, we complement JRC-Acquis with FLORES+~\cite{floresPlus}, a professionally translated benchmark covering all 24 EU languages that is outside the EU legal domain. Using both datasets enables us to verify that our conclusions are robust to potential training data overlap.

\subsubsection{Translation Pipelines}
\label{subsubsec:translation_pipelines}

\paragraph{GPT-4o.} Translations with GPT-4o were performed via the OpenAI API (snapshot \nolinkurl{gpt-4o-2024-08-06}) with \nolinkurl{temperature=0} and a fixed seed to maximize consistency of the outputs. Each request combined a concise instruction and the English source text in a single prompt: ``Translate the following text between double
quotation marks from English to \texttt{\{language\}}:
\texttt{\{text\}}''. The model was queried sequentially, issuing one API call per source text and target language.

\paragraph{eTranslation.} eTranslation required a dedicated infrastructure to handle its asynchronous API and input limit of approximately 2,500 characters. Our pipeline included: (i) segmentation of each policy into linguistically coherent units within the size constraint, (ii) a monitored dispatch system with automatic retries to handle partial failures, and (iii) post-processing scripts to reassemble translations, detect untranslated fragments, and resubmit them until completion. This pipeline ensured that all source documents were fully translated despite intermittent API errors or throttling, providing reproducible and complete outputs across all languages.

\subsubsection{Translation Results}
\label{subsubsec:translation_results}

\begin{figure*}[t]
    \centering
    \includegraphics[width=\linewidth]{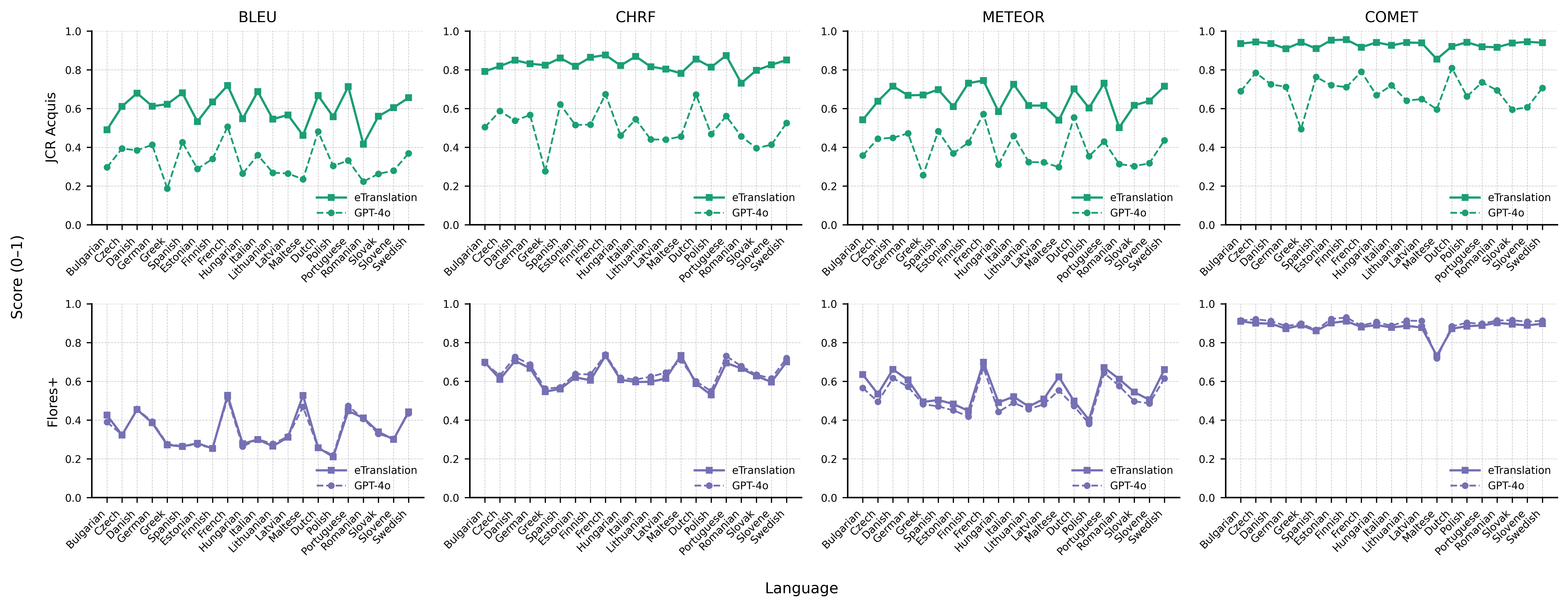}
    \caption{Evaluation of GPT-4o and eTranslation translation quality across 21 EU languages (excluding English, Irish, and Croatian) using BLEU, ChrF, METEOR, and COMET. }
    \Description{A comparative chart displaying the performance of GPT-4o and eTranslation across 21 EU languages using BLEU, ChrF, METEOR, and COMET metrics. The visual trend shows eTranslation consistently achieving higher scores than GPT-4o across all metrics on the JRC-Acquis dataset, with the performance gap being particularly prominent in the COMET metric.}
    \label{fig:translation_results}
\end{figure*}

Figure~\ref{fig:translation_results} summarizes the comparative performance of GPT-4o and eTranslation across the four translation quality metrics on JRC-Acquis and FLORES+ corpora. The analysis covers 21 EU languages, excluding English (used as the source) as well as Irish and Croatian (absent from JRC-Acquis).

On JRC-Acquis, eTranslation consistently outperforms GPT-4o across all metrics. In COMET, eTranslation scores ranged from 0.86 to 0.96, while GPT-4o ranged from 0.49 to 0.81, with the largest differences observed for under-resourced languages such as Greek and Maltese. BLEU, ChrF, and METEOR follow the same trend, showing mean gains of 0.20–0.30 in favor of eTranslation. Figure~\ref{fig:translation_metrics_delta_heatmap} visualizes these cross-lingual gaps, highlighting eTranslation’s stronger advantage for low-resource languages (e.g., Greek, Slovene, Maltese) and its narrower margins for high-resource ones like French, Dutch, and German. These results align with the expectation that eTranslation benefits from domain-specific training on EU legal texts, as discussed in \S\ref{subsec:translation_validation}.

On FLORES+, the pattern changes markedly. Both systems achieve comparable scores, with minor differences in BLEU and ChrF, and only a slight mean advantage for eTranslation in METEOR ($\Delta \approx 0.02$–0.04). COMET likewise shows near-parity, with each system outperforming the other in different languages without any systematic trend. Maltese again represents the lowest-scoring case for both models; yet, their results remain tightly aligned. Overall, FLORES+ demonstrates that outside the EU legal domain, GPT-4o and eTranslation deliver translations of broadly equivalent quality.

\subsection{Dataset Translations}
\label{subsec:dataset_translations}
Given eTranslation's advantage on the legal-domain JRC-Acquis benchmark (\S\ref{subsubsec:translation_results}) and its comparable performance to GPT-4o on FLORES+, we selected eTranslation to generate our corpora. This choice also reflects its stronger alignment with the legal and institutional context of our study: eTranslation is routinely used by EU institutions and public bodies to translate legal and policy materials. This role makes eTranslation a suitable candidate for preserving domain fidelity and linguistic consistency across all official EU languages. Combined with its open availability at no cost, eTranslation is particularly suitable for scaling expert-annotated privacy policy datasets across languages. Following the pipeline described in \S\ref{subsubsec:translation_pipelines}, we translated all OPP-115 and MAPP English-language policies into the 24 official EU languages, producing a linguistically broad derived benchmark for multilingual privacy policy analysis.

\subsection{Legal Expert Sanity Checks}
\label{subsec:manual_validation}

To assess the semantic and terminological fidelity of the translated datasets, we conducted a targeted legal review. A Swiss-trained lawyer with a Master’s degree in law, fluent in English, German, French, and Croatian, examined five representative MAPP policies selected for their annotation density and linguistic complexity. For each policy, translations into French, German, and Croatian were reviewed for semantic accuracy, terminological consistency, and legal clarity, deliberately excluding mere stylistic aspects.

Translation issues were categorized as critical, medium, or low based on their potential to alter the meaning of the original text. The review identified a small number of critical issues in the initial translations, primarily omissions or mistranslations affecting substantive content. Most of these cases corresponded to untranslated fragments in German and Croatian. These issues were automatically corrected through the post-processing step described in \S\ref{subsubsec:translation_pipelines}, after which the corrected translations were re-evaluated by the legal expert. Once the expert confirmed that the mistranslation issues had been resolved, the corrected policy translations were incorporated into the final released dataset. Medium-level issues primarily involved missing section titles or minor inconsistencies in modal expressions, while low-level ones reflected unusual but legally acceptable phrasing. The reviewed deviations did not appear to alter the mapped data-category annotations, although residual issues may remain outside the reviewed sample.

Overall, the legal expert reported that within the reviewed sample, the translations largely preserved the structure and legal semantics of the annotated clauses, with only minor deviations in formatting or terminology.

\section{LLM-based Multilingual Policy Analysis}
\label{sec:multilingual_policy_analysis}

\subsection{Method Design}
\label{subsec:method_design}

To assess whether LLMs can generalize privacy policy analysis across all 24 official EU languages, we extend the method introduced by Rodriguez et al.~\cite{rodriguez2024largeLanguageModelsPrivacyAnalysis} to the multilingual privacy policy benchmark generated in \S\ref{subsec:dataset_translations}. The task remains the same: identifying which categories of personal data are declared as collected in each policy according to the OPP-115 taxonomy\footnote{%
Unlike the original OPP-115 setup, where annotations indicate data collection statements at the segment level, our evaluation operates at the policy level. A privacy policy is considered to declare a given data category if at least one of its annotated segments contains a positive label for that category.} (e.g., contact information, device identifiers, location, financial or health data).

This policy-level classification task requires document-level interpretation rather than isolated segment classification. Unlike a purely lexical matching task, our approach involves interpreting complex linguistic constructions, including negation (e.g., stating that a category is not collected) and conditional statements (e.g., collection occurs only under specific user actions). In addition, the model must preserve conceptual consistency across diverse synonyms and terminologies in 24 languages, so that classifications reflect declared data practices rather than the mere occurrence of specific terms.

Each privacy policy analysis follows a two-message prompting scheme. The first message provides the full policy text in the target language, and the second defines the task together with two few-shot examples in English. These examples---kept identical to the original English setup---illustrate short policy fragments and their corresponding expected outputs. We omit explicit format instructions because OpenAI’s API now supports native structured outputs (\nolinkurl{response\_format="json"}).

All experiments use GPT-4o, replacing GPT-4 Turbo from the prior study by Rodriguez et al.~\cite{rodriguez2024largeLanguageModelsPrivacyAnalysis}, because it provides strong multilingual capabilities, improved API support, and control over generation parameters~\cite{openai2024helloGPT4o}. We set \nolinkurl{temperature=0}, \nolinkurl{top_p=1}, and use a fixed random seed to reduce stochastic variation across runs. GPT-5 and later versions were not used because they do not support configuring parameters such as temperature or random seed.

For each language, the model returns a structured JSON object listing the detected personal data categories. Predictions are aligned with the original English annotations, which we treat as reference labels. Precision, recall, and F1-scores are computed per document and macro-averaged across the dataset. This setup isolates the multilingual dimension by testing whether an English-calibrated method can maintain accuracy across translations without any language-specific adaptation.

\subsection{Cross-Lingual Evaluation}
\label{subsec:cross_lingual_validation}

We first verify that GPT-4o reproduces the English-language results of Rodriguez et al.~\cite{rodriguez2024largeLanguageModelsPrivacyAnalysis}. On the MAPP dataset, the macro-F1 reaches 0.922 ($-1.3$ percentage points (p.p.) relative to ~\cite{rodriguez2024largeLanguageModelsPrivacyAnalysis}); on OPP-115, it reaches 0.939 ($+0.9$ p.p.). These minor deviations are consistent with model updates, confirming that the method behaves stably across model generations.

\begin{figure}[t]
    \centering
    \includegraphics[width=\linewidth]{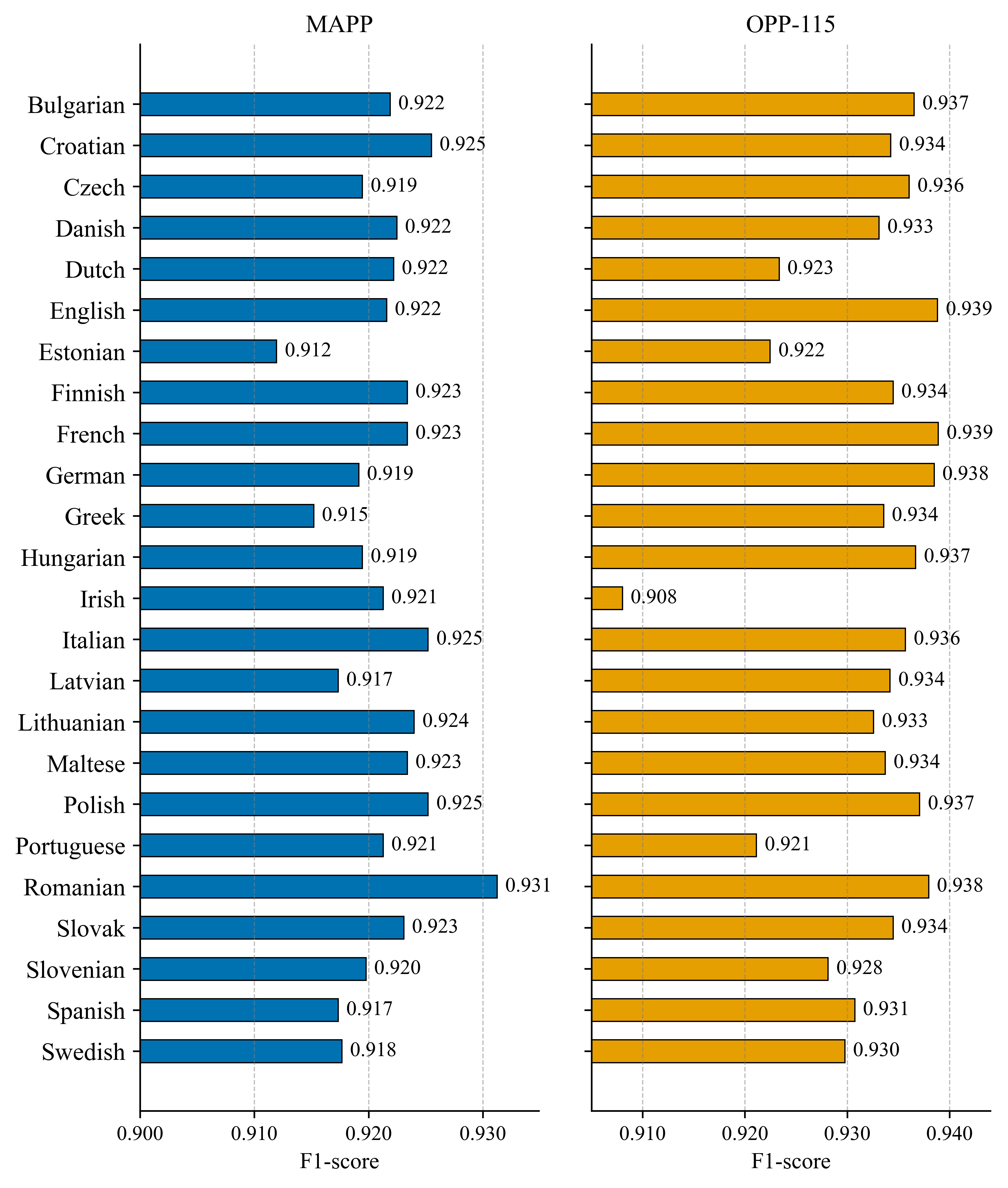}
    \caption{Macro-averaged F1-scores per language for the translated MAPP and OPP-115 datasets. Performance remains uniformly high, showing no systematic language bias and indicating cross-lingual consistency.}
    \Description{A chart displaying macro-averaged F1-scores for 24 EU languages across two datasets, MAPP and OPP-115. The bars are visually flat and tightly clustered between the values of 0.91 and 0.94, illustrating that the classification performance remains stable and uniformly high regardless of the target language.}
    \label{fig:crosslingual_f1}
\end{figure}

We then evaluate cross-lingual classification performance over all 24~EU languages using the translated corpora. Results, summarized in Figure~\ref{fig:crosslingual_f1}, show uniformly high performance: F1-scores range from 0.912  (Estonian) to 0.931 (Romanian) in MAPP and from 0.908 (Irish) to 0.939 (English/French) in OPP-115. Variation across languages remains around three p.p., with no systematic bias favoring any linguistic group. Several languages (e.g., Romanian and Croatian) even slightly outperform English in MAPP, suggesting that, in this benchmark, residual translation noise does not produce observable drops in classification performance.

While partial exposure of OPP-115 or MAPP during pretraining cannot be excluded, the consistency of these results with the high performance observed in the recent independently annotated multilingual corpus from Nenadic et al.~\cite{nenadic2025automatedBoilerplate} provides external support that the results are not solely explained by artifacts of our translated benchmark. These findings show stable performance on translated versions of the source corpora across all EU official languages without the need for fine-tuning or prompt modification, supporting the feasibility of using such methods as part of large-scale multilingual privacy-policy audits.

\subsection{Evaluation with Open-Source Models}
\label{subsec:opensource_validation}

To evaluate the approach beyond proprietary systems, we replicated the multilingual analysis with GPT-OSS-120B, an open-source 120-billion-parameter LLM, using the same configuration and prompting strategy described in \S\ref{subsec:method_design}. Across the 24 official EU languages in OPP-115, GPT-OSS-120B achieved macro-F1 scores between 0.904 and 0.927 ($mean=0.917$), closely matching GPT-4o (0.908--0.939). Variation across languages remained approximately three p.p., with no systematic language bias, suggesting that the approach is not tied to the proprietary model used in the main evaluation.

The open-source configuration also substantially reduces computational cost. Under cloud API pricing at the time of inference, GPT-OSS-120B cost approximately \$0.05 per 1M input tokens and \$0.27 per 1M output tokens, compared with \$2.50 and \$10.00 for GPT-4o, respectively. This 40--50$\times$ reduction suggests that open-source models may support lower-cost multilingual audits for this specific classification task; under comparable benchmark conditions and without reliance on proprietary endpoints.

\section{Application on the Spanish Google Play Store}
\label{sec:spanish_apps}

Our method can support a wide range of real-world applications for privacy policy assessment. In this section, we demonstrate its practical relevance by performing a large-scale audit of Android applications published in the Spanish Google Play Store. This market provides a suitable case study for examining how multilingual privacy policy analysis affects privacy audits, as---in our dataset---public-sector apps mainly publish privacy policies in Spanish, whereas top commercial apps more often provide privacy policies in English. Such linguistic imbalance can complicate cross-sector comparisons when auditing pipelines are limited to English. To evaluate the reliability of privacy policy disclosures, we combine privacy policy analysis with two complementary evidence sources---privacy labels and runtime network traces---allowing us to quantify inconsistencies between declared and observed transmissions of personal data.

\subsection{Experimental Setup}
\label{subsec:experimental_setup}

Our study follows a multi-source design that combines declarative artifacts (policies and labels) with behavioral evidence (network traffic). The corpus includes public sector applications and top-ranked commercial apps to enable a sectoral comparison of disclosure practices. All processing stages run inside a fully automated pipeline that orchestrates app download, installation, execution, traffic capture, and multilingual policy assessment. In total, the pipeline attempted~2{,}611~app executions; the extraction and classification components operate over both Spanish and English policy texts under the same semantic taxonomy (OPP-115), avoiding language-specific adaptations.

The remainder of this section specifies the experiment dataset construction and filtering (\S\ref{subsec:dataset_construction}), the network traffic analysis (\S\ref{subsec:network_analysis}), and the disclosure-consistency analyses (\S\ref{subsec:disclosure_consistency}).

\subsection{Dataset Construction}
\label{subsec:dataset_construction}

The dataset integrates two categories of Android applications distributed through the Spanish Google Play Store:  
(i) apps developed by or for Spanish public institutions, and  
(ii) top-ranked commercial apps across multiple store categories.  
This composition sheds light on the intrinsic differences between public and popular private-sector apps and their respective disclosure practices.

Public-sector applications were identified through a hybrid pipe\-line combining metadata analysis and LLM-based inference. Although Google Play recently introduced a \textit{Government} label, its coverage remains incomplete, and many such apps are still untagged. To achieve broader recall, we issued keyword-based queries targeting institutional domains (e.g., municipalities, regional administrations, or public transport agencies) and retrieved the associated metadata. For apps without the Government tag, we used the DeepSeek-R1 (32B) model as a zero-shot classifier over the app’s title, developer name, and description. Manual validation on a 40-app sample yielded an F1 score of 0.974, indicating high accuracy for large-scale inference. An iterative expansion step incorporated additional apps flagged by the classifier as developed by public entities, yielding 808 government-related applications. The exact search terms used for public-sector app discovery are reported in Appendix~\ref{app:experiment_reproducibility}.

To construct the commercial subset, we retrieved the top 50 apps per Play Store category using a third-party API~\cite{googleplayUnofficialApi}. After deduplication and removal of public-sector entries, the subset comprises 2,556 unique commercial applications. The combined corpus thus includes 3,364 apps gathered from the Spanish Play store.

Of the 3,364 apps, 3,291 were successfully downloaded, and 2,611 could be run on the test devices. Installation failures primarily stemmed from device incompatibility. For each installed app, we collected its privacy label and attempted to retrieve the corresponding privacy policy text from the URL provided in the Play Store metadata, treating this web-hosted document as the disclosure that the developer presents to Google Play users for that app. Retrieved texts were validated using GPT-4o as a binary classifier to determine whether the document constituted a complete privacy policy in Spanish or English. Evaluation on a manually annotated sample (60 per language) yielded an F1 score of 0.956. Policies classified as incomplete or corrupted due to scraping errors or truncated content were excluded.

\begin{figure}[t]
    \centering
    \includegraphics[width=\linewidth]{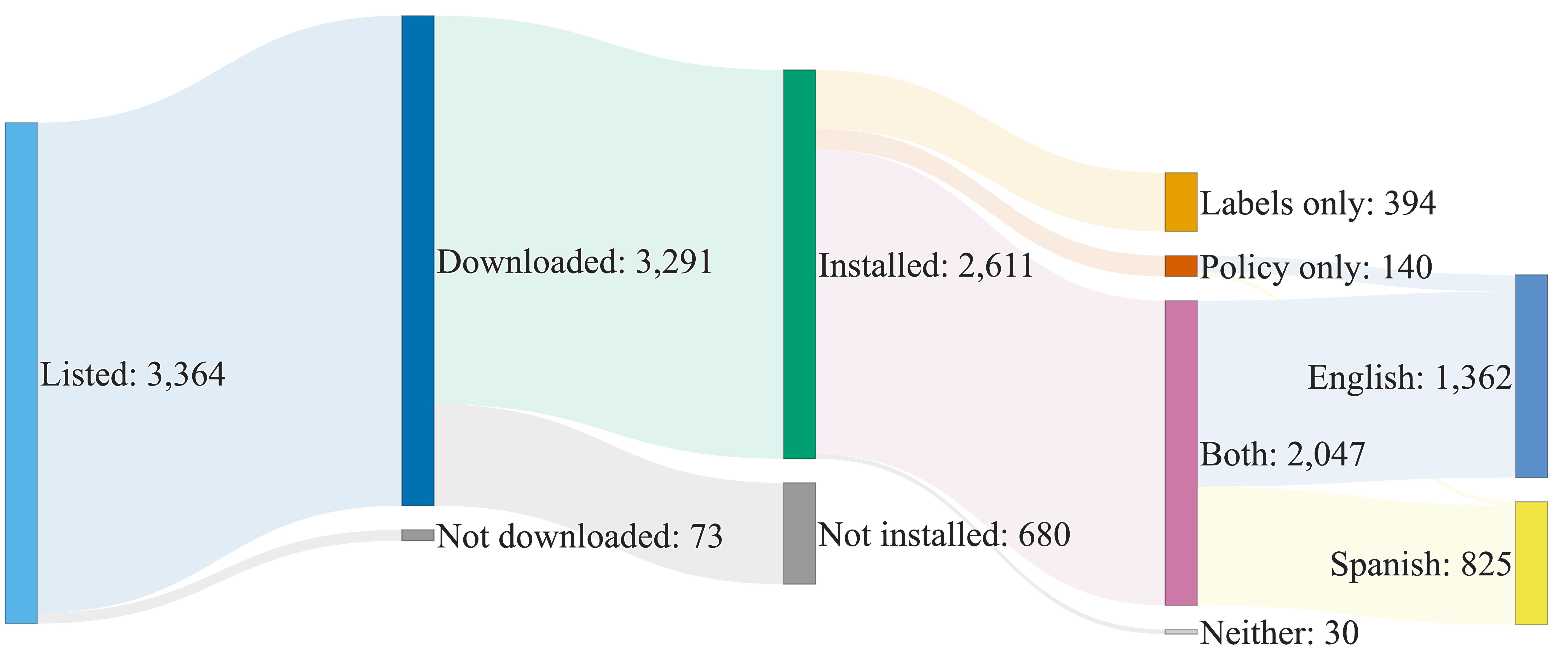}
    \caption{Flow of dataset construction and filtering. The diagram shows the number of applications successfully downloaded, installed, and linked with valid privacy labels or policies, together with the detected privacy policy language.}
    \Description{A Sankey diagram illustrating the data filtering pipeline. It starts on the left with a large block representing 3,364 apps from the Spanish Play Store. The flow progressively branches and narrows down as apps that failed to download or install are filtered out. On the right, it splits into the final subsets showing the number of apps with valid privacy labels and policies, with a further breakdown of policies by their detected language, primarily Spanish and English.}
    \label{fig:dataset_flow}
\end{figure}

After filtering, 2,187 apps contained valid privacy policies (825 in Spanish and 1,362 in English) and 2,441 apps contained valid privacy labels (i.e., Data Safety Section completed by the developer). Among these, 2,047 apps provided both a policy and a label, 394 had only a label, and 140 only a policy. The remaining 30 apps lacking both sources were discarded. Figure~\ref{fig:dataset_flow} summarizes the dataset construction and filtering process.

For the subsequent analyses, we consider two complementary subsets:  
(i) 2,441 runnable apps with valid privacy labels, and  
(ii) 2,187 runnable apps with a valid privacy policy.  
These subsets enable a systematic examination of disclosure consistency by comparing declared and observed data-collection practices (\S\ref{subsec:disclosure_consistency}) and by analyzing cross-sector differences in data practices (\S\ref{sec:cross_sector_practices}).

\subsection{Network Traffic Analysis}
\label{subsec:network_analysis}

\subsubsection{Setup}
\label{subsubsec:network_analysis_setup}

We implemented an on-device dynamic analysis pipeline consistent with prior large-scale mobile-measurement studies~\cite{rodriguez2025sdks, alomar2025effectPlatformPolicies, kollnig2022areIPhonesReallyBetterForPrivacy}. Experiments ran on three rooted Android devices (Xiaomi Redmi 10) with a persistent Frida server and a controlled Wi-Fi environment; Appendix~\ref{app:experiment_reproducibility} provides additional details to support reproducibility. A central orchestration layer coordinated device provisioning, app installation, traffic interception, execution scheduling, and cleanup, minimizing manual intervention.

Before each execution, the orchestrator verified Android Debug Bridge (ADB) connectivity, stopped unrelated background services, checked the interception proxy (\nolinkurl{mitmproxy}~\cite{mitmproxy}), and restored a clean device baseline. Target APKs were installed and their declared runtime permissions programmatically granted to avoid consent dialogs that might alter automated behavior. Network redirection was applied at process level so that only the target app’s traffic was proxied. To inspect encrypted channels without repackaging, Frida~\cite{frida} hooks were injected into Android trust managers and common networking libraries (e.g., \nolinkurl{OkHttp} and \nolinkurl{Conscrypt}) to bypass certificate pinning at runtime.

Each run comprised two controlled phases: a~120-second idle window to capture background or network-initiated activity, followed by a~180-second interaction window driven by \nolinkurl{Android Monkey} with a~150 ms event throttle to emulate user input. At the end of the run, the orchestrator uninstalled the app, terminated residual processes, and reset the device to its baseline state.

All HTTP(S) transactions were logged as structured records (\textit{timestamp}, \textit{ports}, \textit{resolved domain}, \textit{scheme}, \textit{TLS status}, \textit{method}, \textit{URL}, \textit{cookies}, \textit{response code}). Payloads were normalized by a bounded decoding pipeline that sequentially attempts \nolinkurl{gzip/zlib/Brotli} decompression, \nolinkurl{base64/hex} decoding, and URL-form as well as JSON parsing. Personal data transmissions were detected by matching against a curated inventory of identifiers, which included the Google Advertising ID (\texttt{GAID}), Google Service Framework Identifier (\texttt{GSF\_ID}), IP addresses, geolocation coordinates, email addresses, and device-level metadata such as model, build number, and kernel version—across URLs, headers, cookies, and decoded bodies. Encoded or hashed variants of these tokens were also considered to improve recall against lightweight obfuscation.

Destination IPs were geolocated using a commercial IP-to-country service (IPInfo~\cite{ipinfo}) to characterize potential cross-border transfers. The orchestration layer distributed jobs across devices from a central queue and aggregated logs for downstream processing. The resulting dataset provides a systematically captured record of network activity under controlled execution conditions, forming a controlled empirical basis for comparing observed transmissions with reported data collection categories in policies and labels.

\subsubsection{Outbound Data Patterns}
\label{subsubsec:outbound_patterns}

We analyzed the runtime network activity of the 2,611 installed applications to quantify personal data transmissions and characterize their global distribution. Across all executions, 56,810 HTTP(S) flows were intercepted, of which 47,802 (84.1\%) contained at least one element of personal data. While not all transmitted values constitute direct identifiers, many correspond to device-level attributes that can facilitate linkage across sessions or services.

The most frequent elements are summarized in Table~\ref{tab:pii_elements}. \nolinkurl{Device\_Model} and \nolinkurl{Build\_Number} appeared in over 40,000 flows each, indicating widespread transmission of hardware and software metadata. Persistent identifiers such as the \nolinkurl{GAID} and \nolinkurl{GSF\_ID} were found in approximately 2,700 flows each (5.6\% of all personal-data transmissions). Less frequent but more sensitive attributes included fingerprints, device-location coordinates, and network identifiers (e.g., Wi-Fi BSSID or MAC address). Although some of these values may support diagnostic or personalization features, their transmission beyond the device boundary increases the potential for cross-service tracking and re-identification.

\begin{table}[t]
\centering
\caption{Most frequent personal-data elements identified in outbound traffic from 2,611 Android applications. Each entry reports the number of HTTP(S) flows in which the element was detected and its relative share over all flows containing personal data (\(N=47{,}802\)). Multiple elements may co-occur within the same flow.}
\label{tab:pii_elements}
\footnotesize
\begin{tabular}{lrr}
\toprule
\textbf{Data item} & \textbf{Flows} & \textbf{Share of PII-flows} \\
\midrule
\texttt{Device\_Model} & 47{,}443 & 99.2\% \\
\texttt{Build\_Number} & 43{,}193 & 90.4\% \\
\texttt{GSF\_ID} & 2{,}673 & 5.6\% \\
\texttt{GAID} & 2{,}672 & 5.6\% \\
\texttt{Fingerprint} & 1{,}530 & 3.2\% \\
\texttt{Kernel\_Version} & 313 & 0.7\% \\
\texttt{Device\_Location\_Coarse} & 194 & 0.4\% \\
\texttt{Device\_Location\_Precise} & 172 & 0.4\% \\
\texttt{Router\_WiFi\_BSSID} & 147 & 0.3\% \\
\texttt{Build\_Number\_Hash} & 9 & $\approx$0.0\% \\
\texttt{Router\_WiFi\_MAC} & 4 & $\approx$0.0\% \\
\bottomrule
\end{tabular}
\end{table}

Personal-data traffic was geographically concentrated in a limited set of destinations. Most flows terminated in servers in Spain (58.7\%) and the United States (27.8\%), followed by France (4.0\%), Germany (2.9\%), and Ireland (2.5\%). Together, the top 15 countries accounted for over 99\% of all personal-data transmissions. Approximately 69\% of flows were directed to endpoints within the EU, while the remaining 31\% reached non-EEA regions---predominantly the United States, Singapore, India, Hong Kong, and Russia. Although cross-border transfers do not necessarily imply unlawful processing, their prevalence highlights the limited territorial confinement of data flows from applications offered on an EU Play Store.

At the domain level, outbound transmissions exhibited strong concentration. The top 20 recipients accounted for 58\% of all flows containing personal data. Eight belong to the Alphabet ecosystem---\nolinkurl{googleapis.com}, \nolinkurl{googlesyndication.com}, \nolinkurl{gstatic.com}, \nolinkurl{doubleclick.net}, \nolinkurl{google.com}, \nolinkurl{crashlytics.com}, \nolinkurl{googleusercontent.com}, and \nolinkurl{googlevideo.com}---representing roughly 36\% of the total. This dominance reflects the extensive integration of Google-owned SDKs and backend services across both commercial and institutional applications. Other major ad-tech and analytics providers include \nolinkurl{facebook.com}, \nolinkurl{applovin.com}, \nolinkurl{unity3d.com}, \nolinkurl{adjust.com}, and \nolinkurl{appsflyersdk.com}, indicating that mobile data transmission remains heavily concentrated in our dataset.

Regional or app-specific endpoints (e.g., \nolinkurl{dphuesca.es}, \nolinkurl{primor.eu}) collectively contributed only $\approx$2\% of personal data flows. This apparent marginality reflects the inherently fragmented nature of first-party traffic, which is distributed across thousands of first party-owned domains. Conversely, third-party services aggregate data from numerous applications under single endpoints.

\begin{figure}[t]
    \centering
    \includegraphics[width=\linewidth]{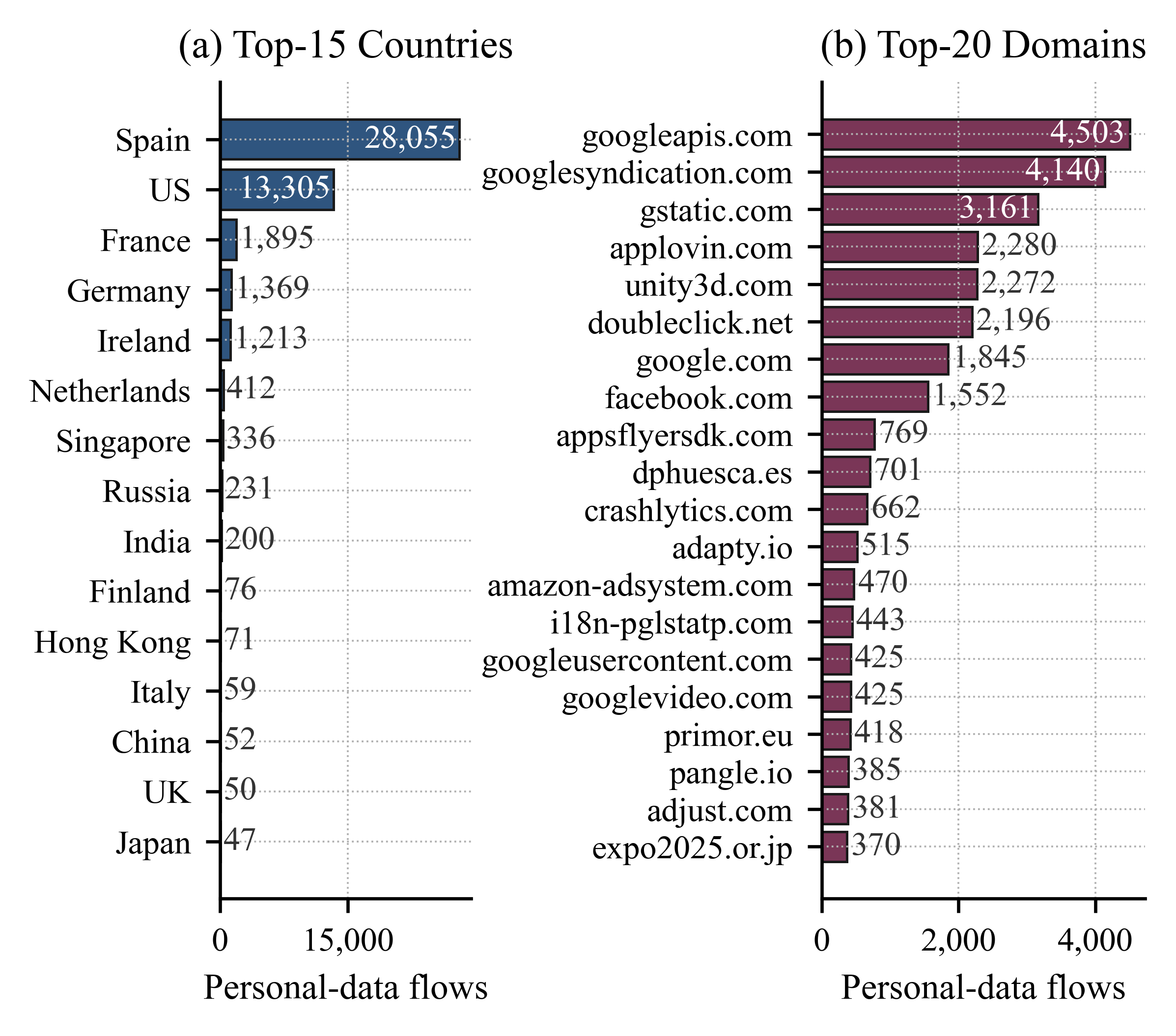}
    \caption{Distribution of outbound personal-data flows observed in analyzed Android applications. Panel (a) shows the top-15 destination countries, with Spain and the United States concentrating the majority of flows. Panel (b) lists the top-20 domains receiving personal data, dominated by Google-owned and advertising-related services.}
    \Description{A two-panel figure showing the destinations of outbound personal-data flows. Panel (a) illustrates the top 15 destination countries, visually highlighting Spain and the United States as receiving the vast majority of the traffic. Panel (b) ranks the top 20 receiving domains, showing a clear dominance of Alphabet-owned services such as googleapis.com and doubleclick.net, followed by other major advertising and analytics platforms.}
    \label{fig:outbound_data}
\end{figure}

\subsection{Disclosure Consistency}
\label{subsec:disclosure_consistency}

We assess disclosure consistency by comparing privacy policies, Google Play’s \emph{Data Safety Section}~\cite{googlePlayDataSafetyHelp}---hereafter \emph{privacy labels} or simply \emph{labels}---and observed network behavior, under a unified taxonomy that semantically aligns declarative and observed data collection categories. Analyses proceed in three stages: (i) Policies vs.\ Behavior, (ii) Labels vs.\ Behavior, and (iii) Labels vs.\ Policies.

\paragraph{Privacy Policy vs Behavior.}
We first assess the alignment between the data collection practices declared in privacy policies and those observed in runtime network transmissions. Among the 2{,}611 analyzed applications, nearly all (2{,}607; 99.8\%) included a privacy policy URL in their Play Store metadata. Our crawler successfully retrieved 2{,}591 documents (99.2\%), of which 2{,}187 (84\%) were validated as genuine privacy policies using the GPT-4o policy-validation classifier described in \S\ref{subsec:dataset_construction}. Within this subset, 467 apps (21.4\%) were developed by public institutions and 1{,}720 (78.6\%) by commercial entities. Dynamic analysis revealed that 1{,}248 applications (57.1\%) transmitted at least one element of personal data.

To operationalize the comparison, we applied the multilingual classification method described in \S\ref{subsec:method_design} to detect, for each validated policy, the OPP-115 personal-data categories declared as collected. Nearly all validated policies (2{,}161; 98.8\%) contained at least one such declaration, confirming that explicit declarations of at least one data category are almost universal among the validated policies in our dataset. 

To align declarative and behavioral evidence, we use the semantic correspondences reported in Table~\ref{tab:policy_label_to_network} (Appendix~\ref{app:semantic_mapping}). These correspondences link each declarative category $c$ to the set $\mathrm{PII}(c)$ of observed data elements that instantiate it (e.g., \emph{Device Model} $\rightarrow$ \emph{Computer information}). The table also includes the equivalent mapping for privacy labels, based on official definitions~\cite{googlePlayDataSafetyHelp}.

Using this mapping, we compared the categories declared in each policy with those inferred from observed transmissions. Let $\mathrm{PII}_a$ denote the set of personal-data elements observed for application $a$, and let $\mathrm{Policy}_a$ be the set of OPP-115 categories declared as collected in its policy. We deem $a$ to exhibit a \emph{disclosure omission} for category $c$ if at least one transmitted element $x \in \mathrm{PII}(c)$ was not declared as collected:
\[
O_{a,c} =
\begin{cases}
1 & \text{if } \exists\, x \in \mathrm{PII}_a : x \in \mathrm{PII}(c) \ \wedge\ c \notin \mathrm{Policy}_a,\\[2pt]
0 & \text{otherwise.}
\end{cases}
\]
The overall omission rate is then:
\[
\mathrm{Omission}_{\mathrm{overall}}
= \frac{\sum_{a} \max_{c} O_{a,c}}{N_{\mathrm{valid}}},
\]
where $N_{\mathrm{valid}}$ denotes the number of applications with validated privacy policies.


Across the 2{,}187 validated policies, 491 applications (22.5\%) exhibited at least one undeclared data-collection practice. Discrepancies were asymmetric: 231 of 467 public-sector apps (49.5\%) displayed omissions, compared with 260 of 1{,}720 popular commercial apps (15.1\%), revealing substantial differences in disclosure completeness. A language-disaggregated view, reported in Appendix~\ref{app:language_disaggregation} (Table~\ref{tab:language_disaggregation}), shows that this asymmetry is also reflected in audit coverage: apps with privacy policies in Spanish account for 825 validated policies, include almost all public-sector apps with valid policies (465 of 467), and contain 310 of the 491 apps with policy omissions. Apps with Spanish and English privacy policies show similar rates of observed personal-data transmission (57.9\% and 56.5\%, respectively), but omission rates differ substantially (37.6\% for Spanish and 13.3\% for English). These results should not be interpreted as a causal effect of policy language, since language is strongly associated with app cohort in our dataset; rather, they show that language coverage determines which app populations and disclosure gaps become measurable. 

At the category level, policy omissions were most frequent for \emph{Computer information} (18.9\%), which includes hardware and system attributes such as \nolinkurl{Device\_Model}, \nolinkurl{Build\_Number}, and \nolinkurl{Kernel\_Version}. Underreporting was also observed for \emph{IP address and device IDs} (4.8\%) and \emph{Cookies and tracking elements} (3.8\%), while \emph{Location} showed the lowest omission rate (0.5\%). Overall, these results indicate a consistent tendency to omit device identifiers and telemetry data, particularly in public-sector applications, which may be problematic from the perspective of the transparency requirements under Articles~12 and~13 GDPR.

\paragraph{Privacy Labels vs Behavior.}
We next assess the consistency between disclosures provided in privacy labels and personal data transmissions observed at runtime. The analysis covered all 2{,}441 applications with labels available, including those explicitly declaring no data collection or sharing. Each label category was semantically aligned with its corresponding set of network-detectable personal data elements according to Table~\ref{tab:policy_label_to_network}, enabling a direct comparison between declared and observed categories. Using the same criterion as above, an application $a$ was considered to omit category $c$ when its traffic included at least one element $x \in \mathrm{PII}(c)$ not declared in its label.

Across the 2{,}441 labeled applications, 188 (7.7\%) exhibited at least one undeclared collection or sharing practice. Omission prevalence was similar between public sector apps (7.1\%) and commercial ones (7.9\%). At the category level, omissions were most frequent for \emph{Device or other IDs} (7.2\%), followed by \emph{Cookies and tracking elements} (1.8\%). \emph{Location} presented the lowest omission rate (0.7\%). Overall, labels appear broadly aligned with observed behavior. However, this outcome should be interpreted cautiously, as the label-to-behavior mapping covers fewer semantic categories than the policy-based comparison, limiting the range of detectable omissions.

\paragraph{Privacy Policy vs Privacy Labels.}
We finally assess the consistency between disclosures presented in privacy labels and those contained in the corresponding privacy policies. Each label category (e.g., \emph{Location}, \emph{Device or other IDs}) was semantically mapped to its most comparable policy category using equivalence ($\equiv$), subset ($\subset$), or superset ($\supset$) relationships, as summarized in Table~\ref{tab:opp115_to_label}.

For each application $a$ and mapping $m$, we define two directional omission indicators:

\begin{align*}
O^{L \to P}_{a,m} &= \mathbf{1}\{\ \mathrm{label}_m(a)=1 \land \mathrm{policy}_m(a)=0\ \},\\
O^{P \to L}_{a,m} &= \mathbf{1}\{\ \mathrm{policy}_m(a)=1 \land \mathrm{label}_m(a)=0\ \}.
\end{align*}

An application is flagged as omitting when at least one direction holds for any mapped category.

To illustrate directionality, consider the mapping \emph{label}~$\subset$~\emph{policy} for \emph{Location}. Privacy labels define \emph{Location} narrowly, distinguishing only between precise and approximate location---roughly corresponding to whether the reported area is smaller or larger than about three square kilometers. In contrast, the OPP-115 \emph{Location} category encompasses broader geographic information such as ZIP code, city, or country-level inferences. Accordingly, omissions are evaluated only in the \emph{Label}~$\rightarrow$~\emph{Policy} direction: if a label declares \emph{Location}, the corresponding policy should also declare it. The inverse direction is not enforced, since a policy may legitimately refer to inferred or coarse geographic data outside the label’s scope.

Our comparison encompassed 2{,}047 applications with both validated policies and labels. Omissions were pervasive: 1{,}451 apps (70.8\%) exhibited at least one directional inconsistency. By direction, 25.4\% of applications showed a \emph{Label} $\to$ \emph{Policy} omission, while 52.4\% displayed the inverse \emph{Policy} $\to$ \emph{Label} pattern. The effect was more pronounced among public sector apps (79.7\%) than among popular commercial ones (68.3\%).

At the level of domain disclosures, discrepancies persisted across all major categories. \emph{Financial information} (modeled as $\equiv$) exhibited a 9.2\% rate of \emph{Label} $\to$ \emph{Policy} omissions. \emph{Location} (mapped as \emph{label} $\subset$ \emph{policy}) produced a 7.5\% omission rate under the same direction. \emph{Contact information} (mapped as \emph{policy} $\subset$ \emph{label}) showed a 3.9\% \emph{Label} $\to$ \emph{Policy} omission rate, alongside additional \emph{Policy} $\to$ \emph{Label} cases. At the data-type level, the strongest divergences arose for identifiers and demographic attributes: \emph{Device or other IDs} $\to$ \emph{IP address and device IDs} affected 7.6\% of apps, while \emph{Other info} $\to$ \emph{Demographic data} reached 7.2\%. Omissions in \emph{Health info} $\to$ \emph{Health/Biometric data} were rare (0.6\%); none were observed for \emph{Race and ethnicity} $\to$ \emph{Demographic data}.

These findings reveal a persistent and asymmetric misalignment between disclosures in privacy policies and labels. Discrepancies are most pronounced in the \emph{Policy} $\to$ \emph{Label} direction, suggesting that developers tend to report a wider range of data categories in their written policies than in privacy labels. From a user perspective, such inconsistencies may obscure the scope of data processing, questioning whether both artifacts jointly deliver the level of transparency envisioned under Articles~12 and~13 of the GDPR.

\newcommand{\PolicyW}{.36\columnwidth}  
\newcommand{\RelW}{.12\columnwidth}     
\newcommand{\LabelW}{.24\columnwidth}   
\newcommand{\MapW}{.18\columnwidth}     

\begin{table}[t]
\centering
\caption{Semantic mapping between OPP-115 data-collection categories and Google Play’s Data Safety Section categories or data types. Relationships indicate whether each label category is equivalent to, contained within, or broader than its policy counterpart.}
\label{tab:opp115_to_label}
\scriptsize
\setlength{\tabcolsep}{2pt}
\renewcommand{\arraystretch}{1.12}
\begin{tabular}{@{}P{\PolicyW} C{\RelW} P{\LabelW} P{\MapW}@{}}
\toprule
\textbf{Policy (OPP-115)} & \textbf{Relation} & \textbf{Label (Data Safety)} & \textbf{Mapping Type} \\
\midrule
Location                & $\supset$ & Location       & \multirow{3}{*}{\textit{Category level}} \\
Contact information     & $\subset$ & Personal info  & \\
Financial info          & $\equiv$  & Financial      & \\
\midrule
IP address and device IDs          & $\supset$ & Device or other IDs & \multirow{4}{*}{\textit{Data type level}} \\
Demographic data                    & $\supset$ & Race and ethnicity  & \\
Demographic data                    & $\supset$ & Other info          & \\
Health, genetic, or biometric data  & $\supset$ & Health info         & \\
\bottomrule
\end{tabular}
\end{table}

\section{Cross-Sector Data Practices} \label{sec:cross_sector_practices}

This section examines how public-sector and commercial applications differ in their data-handling practices. Beyond disclosure completeness, we analyze cross-border transmission patterns, third-party dependency structures, and hosting configurations that shape how personal data elements are transmitted, routed, and disclosed across both cohorts.

\subsection{Cross-Border Flows}
Our measurements reveal systematic contrasts between public-sector and commercial applications in how personal data are transmitted and disclosed. Public-sector apps predominantly communicate with domestic or EU-based servers (87.9\% of transmissions remained within the EEA), whereas popular commercial apps exhibit broader dispersion (67.0\% of transmissions remained within the EEA). Geographical diversity followed a similar pattern: public-sector traffic reached only 11 countries (six within the EEA and five outside), while popular commercial apps sent data to 37 destinations (14 EEA, 23 non-EEA). Although partially attributable to cohort size, these discrepancies point to structurally different integration and hosting practices across ecosystems.

Despite their predominantly EEA-bound profiles, 47 public-sector applications transmitted personal data to endpoints hosted outside the EEA, mainly in the United States and the United Kingdom, including some \emph{.es} domains. Frequent destinations included \nolinkurl{infoplayascanarias.es} (US, 255 flows), \nolinkurl{reskyt.com} (US, 162), and \nolinkurl{traficosevilla.es} (UK, 22), followed by \nolinkurl{cloudflare.com} and \nolinkurl{weatherwidget.org}. Others correspond to auxiliary web infrastructures such as content-delivery networks (e.g., \nolinkurl{gstatic.com}, \nolinkurl{jsdelivr.net}), analytics and crash-reporting frameworks (e.g., \nolinkurl{sentry.io}, \nolinkurl{bugsnag.com}, \nolinkurl{newrelic.com}), or push-notification services (e.g., \nolinkurl{onesignal.com}, \nolinkurl{octopush.me}). While these components often serve legitimate operational purposes, their globally distributed deployments---frequently relying on anycast routing through edge servers~\cite{pascual2024hunterTracingAnycast, pascual2025anycastPrivacyDisaster}---can cause traffic to end up in jurisdictions with heterogeneous data protection standards.

\subsection{Policy Omissions}
When compared with observed network behavior, omissions in privacy policy disclosures were frequent and asymmetric. Among the 39{,}967 network flows ($N=2{,}187$ apps) with valid privacy policies, 10{,}576 (26.5\%) involved at least one category not reflected in the corresponding policy. At the application level, 491 apps (22.5\%) failed to disclose at least one collected data type. Considering only the subset of applications that transmitted personal data (1{,}248), omissions rise to 39.3\%. The imbalance across cohorts is marked: 79.6\% of flows in public-sector apps exhibited omissions, compared with 19.8\% among commercial ones.

Unreported transmissions primarily involved device and system attributes, notably \nolinkurl{Device\_Model} (10{,}096 occurrences) and \nolinkurl{Build\_No} (9{,}502), which together account for more than 95\% of all undeclared items. These attributes appear with comparable frequency across both cohorts and recur in traffic to a wide range of domains, including developer-controlled (first-party) endpoints and external services such as \nolinkurl{googleapis.com}, \nolinkurl{facebook.com}, and \nolinkurl{unity3d.com}. The consistency of these patterns across unrelated domains may suggest that omissions largely stem from shared SDK and runtime dependencies that automatically export telemetry, which may remain opaque to developers~\cite{prybylo2024evaluating}.

In public-sector apps, these attributes were also found in traffic to institutional first-party domains such as \nolinkurl{xunta.gal}, \nolinkurl{emtmadrid.es}, or \nolinkurl{gobiernodecanarias.org}, indicating that even government-operated backends collect system-level identifiers without explicit disclosure in their privacy statements. These results show that incomplete reporting is not confined to monetization contexts but frequently emerges from default engineering practices propagated through common SDKs and service libraries.

\subsection{Third-Party Dependency Patterns}
At a finer level of granularity, cohort-specific destinations reveal distinct third-party dependency structures. Several government-operated applications\allowbreak---\nolinkurl{es.carm.marmenor}, \nolinkurl{org.gobiernodecanarias.ceu.applistareservapas}, and \nolinkurl{es.jcyl.cp.ofar.app}---transmitted undeclared elements such as \nolinkurl{GSF\_ID}, \nolinkurl{Fingerprint}, or \nolinkurl{Device\_Location} (fine or coarse). Network inspection shows that these values rarely target institutional servers; instead, they are routed to third-party infrastructures such as \nolinkurl{googleapis.com} and \nolinkurl{pushwoosh.com}, integrated through mapping, analytics, or notification SDKs. Only isolated cases---such as \nolinkurl{org.gobiernodecanarias.ceu.applista}\allowbreak\nolinkurl{reservapas}, which sent both \nolinkurl{GAID} and \nolinkurl{GSF\_ID} to \nolinkurl{gobiernodecanarias.org}---indicate first-party collection of advertising identifiers. A recurring exception involves \nolinkurl{googleapis.com}, which receives undeclared transmissions from numerous public-sector apps, occasionally including parameters beyond basic device attributes. This pattern highlights the strong coupling of public-sector Android applications with Google’s service frameworks and cloud APIs, even when the European Data Protection Supervisor has warned public institutions about the risks of using cloud-based services~\cite{edps2025microsoft365}.

Commercial applications, in contrast, exhibit a broader and more externally connected data topology. Undeclared transmissions to \nolinkurl{facebook.com} frequently include not only device-related attributes---also observed in public-sector apps---but persistent identifiers such as \nolinkurl{GAID} and \nolinkurl{GSF\_ID}, typically used for advertising and attribution. Both cohorts rely on \nolinkurl{googleapis.com} for backend communication, yet commercial apps more often transmit elements like \nolinkurl{Fingerprint}, \nolinkurl{GAID}, and \nolinkurl{GSF\_ID}, reflecting the integration of SDK modules that extend beyond essential cloud functionality. Comparable undeclared flows were detected toward \nolinkurl{googlesyndication.com}, \nolinkurl{doubleclick.net}, \nolinkurl{unity3d.com}, and \nolinkurl{pangle.io}, which belong to cross-app advertising and analytics infrastructures shared among multiple developers. Overall, our results suggest that public-sector applications tend to centralize data exchanges within controlled backends, whereas commercial apps distribute them across interconnected third-party ecosystems embedded through shared SDKs.

\subsection{Hosting and Transparency Gaps}
Beyond data-collection omissions, hosting configurations expose additional transparency gaps. Several public-sector apps route undeclared identifiers such as \nolinkurl{GAID} or \nolinkurl{GSF\_ID} to first-party domains hosted outside the EEA, including \nolinkurl{traficosevilla.es}, \nolinkurl{infoplayascanarias.es}, and \nolinkurl{cordoba.es}. Although these domains appear governmental, IP geolocation services confirm that their underlying infrastructure resides in non-EU jurisdictions---primarily the United Kingdom and the United States. None of the corresponding privacy policies mention international transfers or specify appropriate safeguards.

This absence leaves users without information on the legal basis or transfer mechanisms governing these flows, undermining transparency and impeding verification of compliance with Articles~13 and~44--49 of the GDPR. The findings suggest that compliance risks in public-sector applications extend beyond SDK telemetry to infrastructure-level dependencies, where outsourcing or external hosting can silently cause personal data to be transferred outside the EU.

\section{Discussion}
\label{sec:discussion}

\paragraph{Mobile privacy measurement in localized ecosystems.}
Prior mobile privacy measurement studies have shown that apps routinely expose personal-data elements in network traffic~\cite{ren2016recon}, often through third-party services used for analytics, advertising, social integration, crash reporting, cloud functionality, or monetization~\cite{vallina2016tracking,Razaghpanah2018Apps}. Large-scale ecosystem studies further show that these flows are concentrated around a limited set of platform and third-party infrastructures~\cite{binns2018third,kollnig2022areIPhonesReallyBetterForPrivacy}. Our measurements are consistent with this literature, but add a localized view of its composition in the Spanish Google Play Store: the dominant pattern is not limited to explicit tracking identifiers or location flows, but includes routine device and system telemetry mediated by platform and service infrastructures. This suggests that apps in this localized market reproduce structural dependencies observed in broader mobile ecosystems, while highlighting how routine telemetry and platform-mediated flows can remain under-described in user-facing disclosures.

\paragraph{Language coverage and public-sector visibility.}
A central contribution of the empirical audit is not the isolated observation that mobile apps transmit personal data, which is well established in prior work, but the ability to shed light on a hitherto less visible app population. Previous large-scale audits have provided broad views of commercial app ecosystems, globally popular apps, and datasets where privacy policies in English dominate~\cite{zimmeck2017automated,zimmeck2019maps,Razaghpanah2018Apps,binns2018third,kollnig2022areIPhonesReallyBetterForPrivacy}. This lens is valuable, but it can underrepresent regional services whose disclosures are written in local languages. In our dataset, this limitation became very concrete: the language-disaggregated analysis shows that public-sector apps overwhelmingly offer privacy policies in Spanish, whereas apps with privacy policies in English are almost entirely popular commercial apps. Multilingual analysis, therefore, changes the audit boundary, determining not only how many policies can be processed, but which parts of the ecosystem become measurable and which transparency gaps can be detected. The resulting comparison shifts the interpretation of mobile transparency gaps from a narrow focus on commercial tracking toward a broader view of how digital services are assembled from external dependencies. This distinction suggests that transparency gaps do not arise from a single tracking-centered model of mobile data collection, but from different dependency structures across app populations. It should not be interpreted as a causal effect of sector alone, since the cohorts differ in purpose, popularity, functionality, development practices, and selection method. Rather, it shows that different app populations exhibit different forms of opacity.

\paragraph{Website policies, mobile apps, and service-level disclosures.}
The study also highlights a methodological nuance in privacy-policy analysis. OPP-115 was originally built from website privacy policies, whereas MAPP and subsequent app-oriented resources focus on policies associated with mobile apps~\cite{wilson2016creation,arora2022corpus,zimmeck2019maps}. This distinction matters because website and mobile contexts differ in their technical affordances, data sources, and platform-specific disclosure requirements. Yet, in deployed mobile ecosystems, the boundary between website and app policies is often porous, as the privacy policy URL listed in an app store frequently points to a web-hosted, service-level document covering websites, accounts, backend processing, analytics, and mobile functionality within the same policy. Evaluating the multilingual classifier on both OPP-115 and MAPP is therefore relevant for our downstream audit, since the method must handle the heterogeneous disclosure artifacts users encounter through app listings, not only policies written exclusively for mobile apps. This also affects how inconsistencies should be interpreted: a broad service-level policy may declare practices not triggered during a specific app execution, whereas an observed app transmission absent from the linked policy indicates that the user-facing disclosure does not cover a practice associated with the app. Our analysis therefore treats privacy policies not as perfect app-specific specifications, but as the transparency artifacts against which observed app behavior can be compared.

\paragraph{From disclosure comparison to continuous oversight.}
Prior work has studied mismatches between app behavior, privacy policies, privacy labels, and internal policy statements~\cite{zimmeck2017automated,zimmeck2019maps,policyLint2019,koch2022keeping,jain2023atlas,ali2024honesty}, and has shown that developer-reported labels are difficult to interpret and maintain~\cite{balash2022longitudinal,khandelwal2024unpacking}. Our results suggest that these inconsistencies should be understood as a traceability problem across the mobile software supply chain. Policies and labels differ in format, level of abstraction, and likely maintenance workflows, while the data flows they describe may originate from SDKs, platform APIs, cloud services, hosting providers, or other dependencies outside the immediate policy-writing process. Improving transparency, therefore, requires mechanisms that connect declared practices to the technical components that generate them, including SDK-level documentation, third-party service inventories, and empirical consistency checks at release time. Platform initiatives such as Apple’s privacy manifests and Google Play’s SDK Index already move in this direction~\cite{applePrivacyManifest,googlePlaySDKIndex}, but our findings suggest that structured declarations should be coupled with empirical verification. Multilingual LLM-based auditing can support this verification layer by bringing non-English policies into the same assessment loop as labels and runtime behavior, extending empirical accountability to localized and public-sector apps that English-centric audits are substantially less likely to cover.

\section{Limitations} \label{sec:limitations}

\subsection{Construct Validity}
The multilingual evaluation benchmark builds upon the expert-annotated OPP-115 and MAPP corpora, which we translated into the 24 official EU languages using the European Commission’s eTranslation system. Although automatic translation can introduce minor semantic drift, translation quality was systematically assessed through multiple complementary metrics and two independent datasets (\S\ref{subsubsec:translation_results}). In addition, a legal expert manually examined a random subset of fifteen translations across diverse language families he is proficient in, providing evidence of semantic and terminological fidelity within the reviewed sample (\S\ref{subsec:manual_validation}). The resulting scores were consistently strong across all languages, and the stability of GPT-4o’s performance (macro-F1 $\approx$ 0.91--0.94) suggests that we did not observe large performance drops attributable to translation artifacts, but we cannot fully disentangle translation quality from downstream classification performance.

A second concern relates to potential exposure of OPP-115 or MAPP during GPT-4o’s pretraining. However, a recent study~\cite{nenadic2025automatedBoilerplate} using an independently constructed multilingual privacy policy dataset reports F1 scores above 0.9 with GPT models, even on unseen data. The close alignment between those external results and our own reduces the concern that the observed performance reflects dataset memorization or benchmark artifacts. Together, the combination of quantitative metrics, manual verification, and cross-literature consistency provides supporting evidence for the construct validity of the multilingual evaluation.

\subsection{Internal Validity}
The dynamic analysis captures direct evidence of personal data transmission by intercepting real network traffic rather than inferring behavior heuristically. Outbound connections from each application were isolated using IPTables rules that redirected only its traffic through the interception proxy, ensuring that all recorded flows originated from the analyzed app. The devices were fully instrumented, enabling maintenance of a verified whitelist of personal data values extracted directly from the device. Matches between these known values and network payloads provide verifiable evidence that personal data were transmitted.

Despite the measures, dynamic testing cannot exhaustively cover all execution paths or user interactions. Certain flows may occur only under extended runtime conditions, authentication, or consent prompts. The reported measurements, therefore, represent a conservative lower bound on data collection activity, as unobserved behavior could reveal additional transmissions. SDKs and backend services may also evolve over time, modifying endpoints or transmission patterns beyond the scope of this snapshot. To attribute destinations, IP-based geolocation was employed, which may be imperfect due to CDN routing or anycast infrastructure. This uncertainty was mitigated through IPinfo---an accurate provider reported in comparative evaluations~\cite{cozar2022reliabilityIpGeolocation}---and by aggregating destinations at the country level to minimize misclassifications when inferring cross-border transfers.

\subsection{External Validity}
The dataset underlying our dynamic analysis comprises two distinct cohorts: (i)~applications developed by or for Spanish public institutions, and (ii)~popular commercial applications distributed through the same store categories. These groups differ structurally in both scope and selection method. The public-sector subset aims for near-complete coverage, encompassing all public-related apps identifiable through the methods described in \S\ref{subsec:dataset_construction}, although few omissions cannot be excluded. This cohort spans a wide popularity spectrum: from widely used regional and national services with millions of installs to narrowly scoped municipal apps with only a few hundred downloads. The commercial subset, in contrast, includes only the most popular apps per category and thus reflects mainstream, high-impact services rather than the full diversity of the private ecosystem. Consequently, our findings characterize disclosure and transmission practices within these two segments but should not be generalized to all applications on the Play Store.

Beyond these cohort-level differences, the dataset includes 2{,}611 Android applications collected from the Spanish Play Store through popular categories sampling and iterative keyword searches. Although designed for broad coverage, the resulting corpus may not encompass all relevant apps, and regional or market-specific variations may exist. Public-sector apps were identified using an automated zero-shot classifier assessing the application's metadata. Validation on a labeled subset yielded high precision and recall, though individual misclassifications cannot be excluded. Given the classifier’s accuracy, these residual errors are unlikely to affect aggregate results.

The analysis represents a single national market and temporal snapshot. Privacy policies, SDK integrations, and data-handling practices evolve over time, and future measurements may yield different prevalence rates. We document the experimental pipeline and configuration to support reproducibility, but dynamic executions may vary slightly due to user interaction, network timing, or SDK behavior. GPT-4o was chosen over newer models (e.g., GPT-5) because it supports fixed temperature and seed parameters, reducing randomness in model outputs. While small deviations across runs or future API versions cannot be excluded, they are unlikely to alter the overall stability of observed trends.


Finally, although the large-scale evaluation used GPT-4o for scalability and API integration, the method does not depend on proprietary infrastructure. The same multilingual pipeline can be reproduced with GPT-OSS-120B (see \S\ref{subsec:opensource_validation}), whose comparable performance (macro-F1 $\approx$ 0.92 across 24 languages) supports open-source deployment.

\section{Conclusions and Future Work} \label{sec:conclusions}
This paper shows that multilingual privacy-policy analysis can be used not only as a classification task, but as an enabling layer for empirical privacy audits in multilingual app ecosystems. Applied to the Spanish Google Play Store, this capability brings public-sector and locally oriented services into the same comparison frame as popular commercial apps. The resulting evidence suggests that many transparency gaps do not represent isolated drafting failures, but rather constitute symptoms of how mobile services are assembled and maintained across technical dependencies and organizational workflows. Across both cohorts, policies and labels often fail to reflect the underlying technical reality. In order to fulfill regulatory transparency requirements, stronger alignment between declarative artifacts and the components that generate data flows remains necessary.

From an infrastructural and regulatory standpoint, three lessons emerge. First, publicly governed translation systems such as the European Commission’s eTranslation can support multilingual processing of legal and policy texts across all official EU languages, but their potential for enabling multilingual transparency analysis in digital services remains largely untapped. Second, automated cross-lingual auditing driven by LLMs can complement existing oversight mechanisms by revealing disclosure gaps early and at large scale, especially when combined with behavioral evidence from runtime analysis. Our open-source model evaluation further suggests a path toward lower-cost and more transparent auditing pipelines, reducing dependence on proprietary endpoints for selected policy-analysis tasks. Third, accountability should extend to the SDK and infrastructure layers, where inherited data collection behaviors need to be documented and disclosed consistently across integrated components.

Future research should build on these findings to conduct comparative studies across regions, languages, and sectors, as well as to extend auditing toward other privacy dimensions such as data sharing, retention, and purpose limitation. Expanding multilingual evaluation beyond a single national market would illuminate how transparency obligations are interpreted and implemented across jurisdictions, advancing toward privacy compliance that is empirically grounded at ecosystem scale. The same methodological foundations could support auditing in other legal domains, from consumer protection to financial contracts, demonstrating the broader potential of LLM-based auditing for regulatory oversight.



\normalsize
\bibliography{references}


\appendix
\section{Appendix}

\subsection{Cross-lingual Quality Differences}
\label{app:crosslingual_quality}

\begin{figure}[H]
\centering
\includegraphics[width=\linewidth]{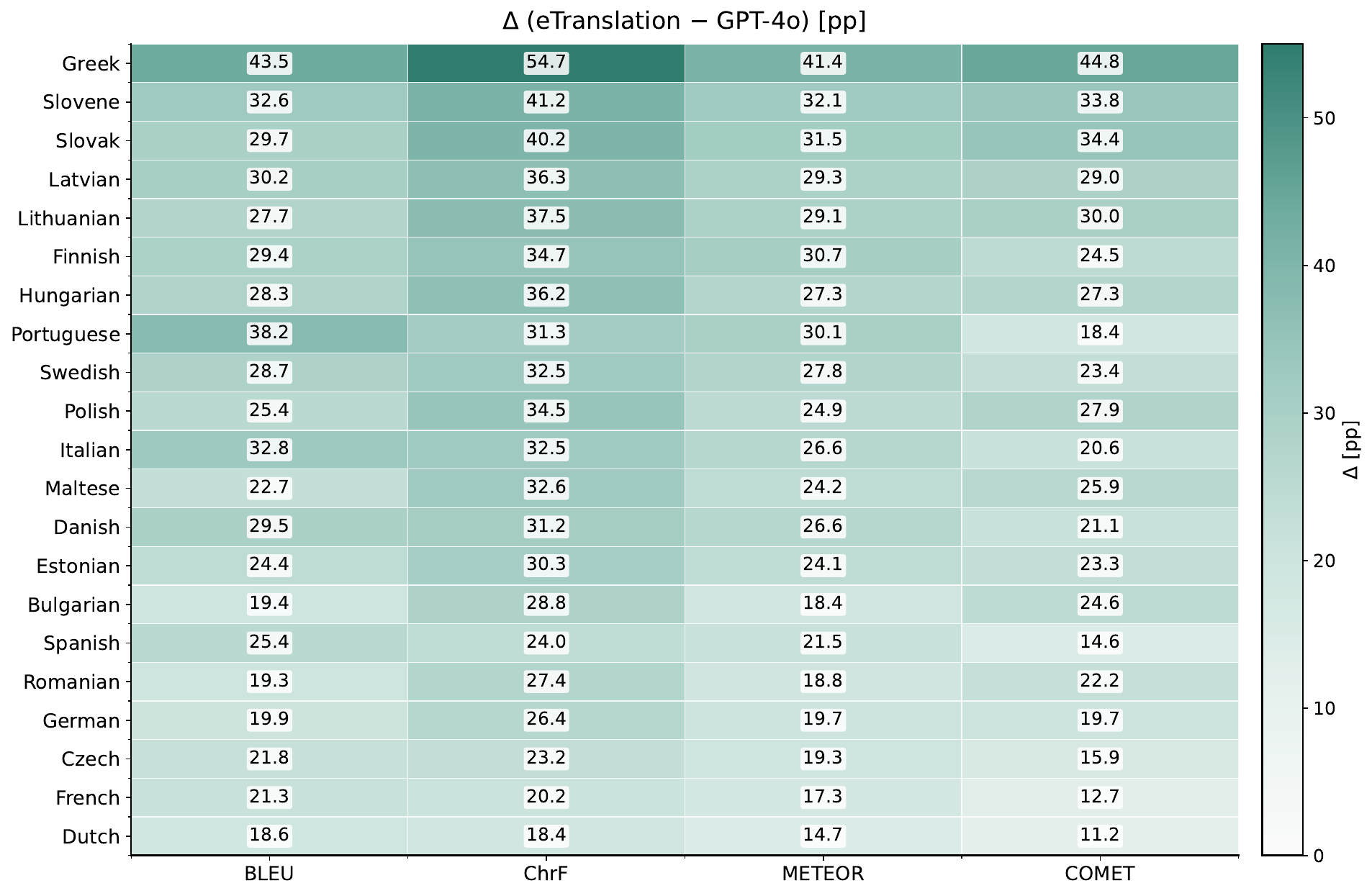}
\captionof{figure}{Cross-lingual differences in translation quality between eTranslation and GPT-4o across the 24 official EU languages using JRC-Acquis as reference. Values represent average score deltas (eTranslation~--~GPT-4o) in percentage points for BLEU, ChrF, METEOR, and COMET, sorted by overall gap. Larger positive values indicate a greater relative advantage of \texttt{eTranslation} for the corresponding language.}
\label{fig:translation_metrics_delta_heatmap}
\Description{A heatmap visualizing the performance gap between eTranslation and GPT-4o across 24 languages and 4 translation metrics. The color intensities show that eTranslation has a much stronger relative advantage in lower-resource languages like Greek, Slovene, and Maltese, whereas the color gradient becomes very light for high-resource languages like French and German, indicating a narrower performance gap.}
\end{figure}

Appendix Figure~\ref{fig:translation_metrics_delta_heatmap} reports the cross-lingual differences in translation quality between eTranslation and GPT-4o on the JRC-Acquis benchmark.

\subsection{Experiment Reproducibility}
\label{app:experiment_reproducibility}

To compile the dataset of applications, we developed a custom collection script targeting the Spanish Google Play Store. Public-sector candidates were retrieved through keyword-based searches using Google Play URLs of the form \url{https://play.google.com/store/search?q=<KEYWORD>&c=apps}. The search terms were selected to cover Spanish public services, administrative entities, transport services, health services, authentication systems, and major municipalities. The full set of keywords was: `transporte', `ayuntamiento', `comunidad', `gobierno', `salud', `taxi', `autobus', `metro', `madrid', `barcelona', `valencia', `bilbao', `clave', `FNMT', `museo', `tramites', `oficial', `eGob', `ciudadano\allowbreak+\allowbreak digital', `eadministracion', `sevilla', `zaragoza', `malaga', `murcia', `palma\allowbreak+\allowbreak de\allowbreak+\allowbreak mallorca', `las\allowbreak+\allowbreak palmas\allowbreak+\allowbreak de\allowbreak+\allowbreak gran\allowbreak+\allowbreak canaria', `alicante', `cordoba', `valladolid', `vigo', `gijon', `ciudadano', `nacional', `ministerio', and `ministerio\allowbreak+\allowbreak para\allowbreak+\allowbreak transf\allowbreak+\allowbreak digital\allowbreak+\allowbreak y\allowbreak+\allowbreak funcion\allowbreak+\allowbreak publica'. Commercial applications were collected separately as the top-ranked applications per Google Play category, as described in Section~\ref{subsec:dataset_construction}. Duplicates and applications classified as public-sector apps were removed from the commercial subset.

The dynamic analysis was performed using three Xiaomi Redmi 10 smartphones. Each device has an octa-core processor of up to 2.0GHz, 4GB of RAM, and 64GB of internal storage. All devices ran Android 11 (build RP1A.200720.011) with MIUI version 12.5.1.0(RKUEUHG). To enable traffic inspection, the devices were rooted using Magisk. We installed Cert Fixer version 1.1 by pwnlogs to move Certificate Authority (CA) certificates from the user store to the system partition, and MagiskFrida version 17.4.0-1 by ViRb3 and enovella, which automatically launches a frida-server of the same version on device startup.

Each application execution followed the protocol described in Section~\ref{subsubsec:network_analysis_setup}: an idle phase followed by an interaction phase. To simulate interactions, we used the \texttt{monkey} tool included in Android Debug Bridge (ADB), together with a custom script that performs swipes and taps in pseudorandom screen regions. The script was intentionally not configured to fill in app placeholders such as email addresses, usernames, passwords, or other user-provided identifiers.

Applications that failed installation, could not be executed on the test devices, or detected the rooted environment and refused to run were excluded from the corresponding analyses. After each run, the app was uninstalled, residual processes were terminated, and the device was restored to the baseline state before processing the next application.

\subsection{Data Category Mapping and Semantic Alignment}
\label{app:semantic_mapping}

This appendix reports the semantic alignment between declarative data-collection categories and their observable counterparts in network traffic. Table~\ref{tab:policy_label_to_network} defines the mapping used to identify disclosure omissions and inconsistencies across privacy policies, labels, and behavioral evidence.

\begin{table*}[t]
\centering
\caption{Semantic mapping between declarative data-collection categories and their observable counterparts in network traffic.}
\label{tab:policy_label_to_network}
\scriptsize
\begin{threeparttable}
\setlength{\tabcolsep}{6pt}
\renewcommand{\arraystretch}{1.12}
\begin{tabularx}{\textwidth}{@{}l l Y T@{}}
\toprule
\textbf{Domain} & \textbf{Category} & \textbf{Definition / Scope} & \textbf{Associated PII} \\
\midrule
\multirow[t]{4}{*}{\textbf{Policy domain}}
& Location &
Geo-location information about a user’s position, regardless of granularity (e.g., exact location, ZIP code, city level). &
\piilist{\nolinkurl{Device\_location}, \nolinkurl{Device\_location\_coarse}, \nolinkurl{Router\_WiFi\_BSSID}, \nolinkurl{Router\_WiFi\_MAC}} \\
\addlinespace[2pt]
& IP address and device IDs &
Permanent (e.g., device IDs, MAC address) or temporary (e.g., IP address) identifiers needed to establish a connection for the current session. &
\piilist{\nolinkurl{GSF\_ID}, \nolinkurl{GAID}, \nolinkurl{Fingerprint}} \\
\addlinespace[2pt]
& Cookies and tracking elements &
Locally stored identifiers (e.g., cookies, beacons) commonly used to uniquely identify users but not essential to establish a connection. &
\piilist{\nolinkurl{GAID}}\tnote{*} \\
\addlinespace[2pt]
& Computer information &
Type of operating system or browser used, or similar device-level information. &
\piilist{\nolinkurl{Device\_Model}, \nolinkurl{Build\_No}, \nolinkurl{Kernel\_Version}} \\
\midrule
\multirow[t]{2}{*}{\textbf{Label domain}}
& Device or other IDs &
Identifiers that relate to an individual device, browser, or app instance (e.g., IMEI, MAC address, Advertising ID, Firebase installation ID). &
\piilist{\nolinkurl{GSF\_ID}, \nolinkurl{GAID}, \nolinkurl{Fingerprint}} \\
\addlinespace[2pt]
& Location &
User or device physical location (precise \nolinkurl{ACCESS\_FINE\_LOCATION} or approximate \nolinkurl{ACCESS\_COARSE\_LOCATION}), or derived from network information. &
\piilist{\nolinkurl{Device\_Location}, \nolinkurl{Device\_Location\_Coarse}, \nolinkurl{Router\_WiFi\_BSSID}, \nolinkurl{Router\_WiFi\_MAC}} \\
\bottomrule
\end{tabularx}
\begin{tablenotes}\footnotesize
\item[*] While the Google Advertising ID is formally categorized as a device identifier, it is functionally equivalent to a tracking element enabling cross-app linkage and profiling, and was therefore included under this class.
\end{tablenotes}
\end{threeparttable}
\end{table*}

\subsection{Policy Language and Audit Coverage}
\label{app:language_disaggregation}

Table~\ref{tab:language_disaggregation} provides the language-disaggregated results underlying the policy--behavior analysis. We use this comparison descriptively, since the language of privacy policies is strongly associated with app cohort in our dataset; it therefore characterizes which applications become measurable under multilingual auditing rather than estimating a causal effect of language.

\begin{center}
\begin{threeparttable}
\captionof{table}{Distribution of validated policies, observed personal-data transmissions, and policy omissions by privacy policy language.}
\label{tab:language_disaggregation}
\scriptsize
\setlength{\tabcolsep}{1.2pt}
\renewcommand{\arraystretch}{1.18}
\begin{tabular}{@{}
>{\raggedright\arraybackslash}m{0.16\columnwidth}
>{\centering\arraybackslash}m{0.07\columnwidth}
>{\centering\arraybackslash}m{0.14\columnwidth}
>{\centering\arraybackslash}m{0.19\columnwidth}
>{\centering\arraybackslash}m{0.21\columnwidth}
>{\centering\arraybackslash}m{0.15\columnwidth}
@{}}
\toprule
\textbf{Language} &
\textbf{Apps} &
\textbf{Public} &
\textbf{Commercial} &
\textbf{Apps transmitting personal data} &
\textbf{Apps with policy omission} \\
\midrule
Spanish & 825 & 465 (56.4\%) & 360 (43.6\%) & 478 (57.9\%) & 310 (37.6\%) \\
English & 1{,}362 & 2 (0.1\%) & 1{,}360 (99.9\%) & 770 (56.5\%) & 181 (13.3\%) \\
\bottomrule
\end{tabular}

\begin{tablenotes}[flushleft]
\scriptsize
\item[] \emph{Note:} Percentages are computed within each policy-language group. Commercial denotes apps in the popular commercial cohort.
\end{tablenotes}
\end{threeparttable}
\end{center}
\end{document}